# The Zenon effect in Quantum Mechanics

Nikos Bagis
Department of Informatics, Aristotele University of Thessaloniki
Thessaloniki 54006 Greece
bagkis@hotmail.com

**Abstract**
The basis of the so-called Zenon effect in Quantum Mechanics, is the limiting behavior of the unitary solution of Schroedinger`s differential equation, under repeated measurements. We examine the limit of a sequence of operators composed by a usual operator and an arbitrary projection operator.

**Introduction**

Recall that the Schrödinger equation describes the total energy of a particle in terms of potential and dynamical energy:

$$i\hbar \frac{\partial}{\partial t}\Psi = -\frac{\hbar^2}{2m}\nabla^2\Psi + V(x)\Psi,$$

where $\hbar$ is Planks constant, $m$ is the mass of particle, $V(x)$ is the potential, $\Psi(x,t) \in L^2(\mathbb{R}^n \times \mathbb{R}^+)$, $x$ is the position of the particle at time $t$. The evolution of the quantum system is expressed by $e^{-itH/\hbar}$, where $H = -\frac{\hbar^2}{2m}\nabla^2 + V(x)$ is the Hamiltonian of the quantum system.

Frequent measurement can slow the time evolution of a quantum system, hindering transitions to states different from the initial one (A.Beskow and J.Nilsson, 1967 and B.Misra and E.C.G Sudarshan, 1977).

This phenomenon is known as the quantum Zeno effect. One can formulates the Quantum Zenon Effect as follows:

Consider a quantum system whose states belong to the Hilbert space $L^2$ and whose evolution is described by the unitary operator $U_t = \exp(-iHt/\hbar)$, where $H$ is a time-independent lower-bounded Hamiltonian, and $E_M L^2 = L^2_M$ the subspace defined by it. The Zeno effect problem can be focused on the limit operator

$$W_T^N = \left(E_M U_{T/N} E_M\right)^N, \text{ as } N\to\infty, N=1,2,3,...$$

If this limit exists, then the limit operators $W$ (say) form a one-parameter semigroup, and the probability to find the quantum system in $L^2_M$ is 1 (according to Misra and Sudarshan). In other words if the limit exists the particle remains in $L^2_M$.

In this work we study a model based by an infinite sequence: $W_T^N = \left(E_M U_{T/N} E_M\right)^N$, $N = 1,2,…$, of operators on $L^2$, where $T \in \mathbb{R}^+$, $E_M$ is a projection operator and $U_s$ is a unitary operator on $L^2$

Notice that this problem is rather obvious on certain cases of projection operators $E_M$ commuting with the translation operator, we give the following examples:

**Examples**

**a)** Let $E_M$ be frequencies truncation operator: $f \to E_M f$, who sends an element of $L^2(\mathbb{R})$ to $L^2_M$, or for simplicity $M$, $E_M f(x) = \frac{1}{2\pi}\int_{-k}^{k} f^\wedge(\gamma)e^{i\gamma x}d\gamma, x \in \mathbb{R}$, for some $k > 0$, we have

$((U_T f)(t) = f(t+T)$ is the usual translation operator)

$$U_T E_M f(x) = \frac{1}{2\pi}\int_{-k}^{k} f^\wedge(\gamma)e^{i\gamma(x-T)}d\gamma = \frac{1}{2\pi}\int_{-k}^{k} f^\wedge(\gamma)e^{i\gamma x}e^{-i\gamma T}d\gamma = E_M U_T f(x), t, x \in \mathbb{R}.$$

Thus $E_M U_T E_M = U_T E_M$ and so: $W_T^N = \prod_{n=1}^{N}\left(E_M U_{a_n(T)} E_M\right) = U_{\sum_{n=1}^{N} a_n(T)} E_M$

**b)** Let $E_M f(x) = \sum_{n=-N}^{N} f^\wedge(n)e^{inx}$, be the projection operator on the space of all trigonometrical polynomial of degree $N$, $N = 2, 3\ldots$ as in (a), it is not difficult to see that $E_T^N = \prod_{n=1}^{N}\left(E_M U_{a_n(T)} E_M\right) = U_{\sum_{n=1}^{N} a_n(T)} E_M$.

## Section 1
## The general case of Zeno effect

**Definition 1.**
Let $A \subseteq \mathbb{R}$ and $U_s : L^2(A) \to L^2(A)$ be a bounded operator such that:
**i)** $\lim_{s \to 0} U_s f(x) = f(x)$

**ii)** The derivatives of $U_s$: $\left.\frac{d^r}{ds^r}U_s\right|_{s=0}$, for $r = 1, 2$, around $s = 0$ exists and they are bounded operators on $L^2(A)$.

Note that we can take as operator $U_s f(x) = \int_A f(t)K_s(x,t)dt$, satisfying the conditions (i), (ii) of Definition 1 which is a very general case. Also we did not restrict our self's that $U_s$ is unitary. This requirement we will get latter.

**Definition 2.**
Let $M$ is an arbitrary subspace of $L^2(A)$ and let $\{e_k\}_{k\in\mathbb{N}}$ is an orthonormal base of $M$.
**i)** We call $R_{k,m}(s) := \langle U_s e_k | e_m \rangle$, for every $k, m$ in $\mathbb{N}$.
**ii)** $E_M$ is the projection operator on $M$.

**Definition 3.**

We call Propagator the family of matrices $G_N$, $N = 1, 2, ...$ who have elements:

$$G_{k,k_N}\left(\tfrac{T}{N}\right) := \sum_{k_1=0}^{\infty} R_{k,k_1}\left(\tfrac{T}{N}\right) \sum_{k_2=0}^{\infty} R_{k_1,k_2}\left(\tfrac{T}{N}\right) \cdot ... \cdot \sum_{k_{N-1}=0}^{\infty} R_{k_{N-2},k_{N-1}}\left(\tfrac{T}{N}\right) R_{k_{N-1},k_N}\left(\tfrac{T}{N}\right)$$

**Definition 4.**
$B(s)$ will be a linear operator $M \to M$, such that

$$\sum_{k=-\infty}^{\infty} \langle f | e_k \rangle e_k(x) \to \sum_{k=-\infty}^{\infty} \langle f | e_k \rangle \sum_{m=0}^{\infty} R_{k,m}(s) e_m(x)$$

where $\{e_k(x)\}_{k \in \mathbb{Z}}$ is a base of $M$.

**Lemma 1.**
We will show that under the conditions taken in Definitions 1, 2, that
i) $B(s) = E_M U_s$ in $M$

ii) $\dfrac{d^r}{ds^r} B(s) = E_M \left( \dfrac{d^r}{ds^r} U_s \right)$, $r = 1, 2$

iii) It is $B^{(r)}(0)(e_k(x)) = \sum_{m=0}^{\infty} \left. \dfrac{d^r R_{k,m}(s)}{ds^r} \right|_{s=0} \cdot e_m(x)$, for $r = 0, 1, 2$ and if $k \in \mathbb{N}$:

$\sum_{m=0}^{\infty} |R_{k,m}(s)|^2$, $\sum_{m=0}^{\infty} |R'_{k,m}(s)|^2$, $\sum_{m=0}^{\infty} |R''_{k,m}(s)|^2 < \infty$ around $s = 0$.

**Proof.**

Let $f$ be on in $M$ then $f(x) = \sum_{k=-\infty}^{\infty} \langle f | e_k \rangle e_k(x)$

$E_M U_s f = \sum_{m=0}^{\infty} \langle E_M U_s f | e_m \rangle e_m(x) =$

$= \sum_{m=0}^{\infty} \langle U_s f | e_m \rangle e_m = \sum_{m=0}^{\infty} \langle U_s \sum_{l=0}^{\infty} \langle E_M f | e_l \rangle e_l | e_m \rangle e_m =$

$= \sum_{m=0}^{\infty} \langle \sum_{l=0}^{\infty} \langle f | e_l \rangle U_s e_l | e_m \rangle e_m = \sum_{l=0}^{\infty} \langle f | e_l \rangle \sum_{m=0}^{\infty} \langle U_s e_l | e_m \rangle e_m =$

$= \sum_{l=0}^{\infty} \langle f | e_l \rangle \sum_{m=0}^{\infty} R_{l,m}(s) e_m = B(s) E_M f$

The operator $B$ is clearly bounded as the product of two bounded operators.
We next show that $\sum_{m=0}^{\infty} |R_{k,m}(s)|^2 < \infty$, for every $k \in \mathbb{N}$.

$\sum_{m=0}^{\infty} |R_{k,m}(s)|^2 = \sum_{m=0}^{\infty} |\langle U_s e_k | e_m \rangle|^2 = \|U_s e_k\|_2^2 < \infty$

The same we can do with $B'(0) = E_M \left. \dfrac{dU_s}{ds} \right|_{s=0}$ and $B''(0) = E_M \left. \dfrac{d^2 U_s}{ds^2} \right|_{s=0}$

**Lemma 2.**

Let $W_T^{(N)} := (E_M U_{T/N} E_M)^N$, $N \in \mathbb{N}$. Then the following relation holds:

$$W_T^{(N)} f(x) = \sum_{k=0}^{\infty} \langle f | e_k \rangle \sum_{k_N=0}^{\infty} G_{k,k_N}\left(\tfrac{T}{N}\right) e_{k_N}(x)$$

**Proof.**
Easy

**Note.** We use the notation $e^B$ for the transform $\sum_{k=0}^{\infty} \dfrac{B^k}{k!}$, whenever $B$ is a bounded operator (See **[G]**).

**Lemma 3.**
For the transformation $B(s)$ we have

i) $B\left(\tfrac{T}{N}\right)^N e_k(x) = \sum_{k_N=0}^{\infty} G_{k,k_N}\left(\tfrac{T}{N}\right) e_{k_N}(x)$.

ii) $\lim_{N \to \infty} \left\| \left( I + \dfrac{T}{N} B'(0) \right)^N - e^{TB'(0)} \right\| = 0$.

**Proof.**
i) We have:
$$B(s)^2 e_k(x) = \sum_{m=0}^{\infty} R_{k,m}(s) B(s) e_m(x) = \sum_{m=0}^{\infty} R_{k,m}(s) \sum_{n=0}^{\infty} R_{m,n}(s) e_n(x).$$
In general:
$$B(s)^N e_k(x) = \sum_{k_1=0}^{\infty} R_{k,k_1}\left(\tfrac{T}{N}\right) \sum_{k_2=0}^{\infty} R_{k_1,k_2}\left(\tfrac{T}{N}\right) \cdot \ldots \cdot \sum_{k_N=0}^{\infty} R_{k_{N-1},k_N}\left(\tfrac{T}{N}\right) e_{k_N}(x)$$
and the result follows from definition 3.

We will show that: $\left\| \left(1 + \dfrac{B'(0)T}{N}\right)^N - e^{B'(0)T} \right\| = O\left(\tfrac{1}{N}\right)$, $\mathbb{N} \ni N \to \infty$.

For to complete our purpose we need an asymptotic formula (see **[A,S]** pg 257):
Let $\Gamma$ be the Gamma function. For $a$ real and positive and $N \in \mathbb{N}$, the following asymptotic formula is valid:
$$\Gamma(aN + b) = \sqrt{2\pi} e^{-aN} (aN)^{aN+b-1/2} \left(1 + O\left(\tfrac{1}{N}\right)\right), \quad N \to \infty.$$
Now:
$$\left\| \left(I + \tfrac{TB'(0)}{N}\right)^N - e^{TB'(0)} \right\| = \left\| \sum_{k=0}^{N} \binom{N}{k} \left(\dfrac{TB'(0)}{N}\right)^k - \sum_{k=0}^{\infty} \dfrac{T^k (B'(0))^k}{k!} \right\| =$$

$$\left\| \sum_{k=0}^{N} \dfrac{T^k (B'(0))^k}{k!} \left( \dfrac{N!}{(N-k)! N^k} - 1 \right) - \sum_{k=N+1}^{\infty} \dfrac{(B'(0))^k T^k}{k!} \right\| \le$$

$$\left\|\sum_{k=0}^{N}\frac{(B'(0))^k T^k}{k!}\left(\frac{N!}{(N-k)!N^k}-1\right)\right\| + \left\|\sum_{k=N+1}^{\infty}\frac{(B'(0))^k T^k}{k!}\right\| \leq$$

$$\sum_{k=0}^{N}\frac{\|B'(0)\|^k T^K}{k!}\left|\frac{N!}{(N-k)!N^k}-1\right| + \sum_{k=N+1}^{\infty}\frac{\|B'(0)\|^k T^k}{k!} =$$

$$\sum_{k=0}^{N}\frac{\|B'(0)\|^k T^k}{k!}\left|\frac{\Gamma(N+1)}{\Gamma(N-k+1)N^k}-1\right| + \sum_{k=N+1}^{\infty}\frac{\|B'(0)\|^k T^k}{k!} \leq$$

$$\leq \sum_{k=0}^{N}\frac{\|B'(0)\|^k T^k}{k!}\left|\frac{\sqrt{2\pi}e^{-N}N^{N+1-1/2}(1+O(\tfrac{1}{N}))}{\sqrt{2\pi}e^{-N}N^{N-k+1-1/2}(1+O(\tfrac{1}{N}))N^k}-1\right| + \sum_{m=0}^{\infty}\frac{\|B'(0)T\|^{m+N+1}}{\Gamma(m+N+2)} =$$

$$= \sum_{k=0}^{N}\frac{\|B'(0)\|^k T^k}{k!}O(\tfrac{1}{N}) + \|B'(0)T\|^{N+1}\sum_{m=0}^{\infty}\frac{\|B'(0)T\|^m}{\sqrt{2\pi}e^{-N}N^{N+m+2-1/2}} = O\left(\frac{e^{\|B'(0)\|T}}{N}\right) = O(\tfrac{1}{N})$$

**Lemma 4.**
For the operator $B(h)$ we have, for $h$, $s$ small, that there exists a number $\xi \in [0,s]$ such that:

$$B(h) = I + hB'(0) + \frac{h^2}{2}B''(\xi)$$

Where $If(x) = f(x)$.

**Proof.**
$$\lim_{N\to\infty} R_{k,m}\left(\tfrac{T}{N}\right) = \lim_{N\to\infty}\langle U_{T/N}(e_k(x)) | e_m(x)\rangle = \langle e_k(x) | e_m(x)\rangle = \delta_{k,m}$$

From Taylor's expansion theorem we get that:

$R_{k,m}(h) = \delta_{k,m} + hR'_{k,m}(0) + \frac{h^2}{2}R''_{k,m}(\xi)$, where the $R''_{k,m}(\xi)$ are bounded for all $k$, $m$.

So from the definition of $B(s)$ we get the lemma.

**Lemma 5.**
$$\lim_{N\to\infty}\left(B\left(\tfrac{T}{N}\right)\right)^N = e^{TB'(0)}$$

**Proof.**
We have $\left(B\left(\tfrac{T}{N}\right)\right)^N = \left(I + \tfrac{T}{N}B'(0) + \tfrac{1}{2}\left(\tfrac{T}{N}\right)^2 B''(\xi)\right)^N =$
$\left(I + \tfrac{T}{N}B'(0)\right)^N + O(\tfrac{1}{N}) \to e^{TB'(0)}$, as $N \to \infty$.

**The Main Theorem.**
The limit operator $W_T f(x) = \lim_{N\to\infty}(E_M U_{T/N} E_M)^N f(x)$ exists when

**i)** $\lim_{s\to 0} U_s f(x) = f(x)$

**ii)** The derivatives of $U_s$: $\left.\dfrac{d^r}{ds^r}U_s\right|_{s=0}$, for $r = 1,2$, around $s = 0$ exists and they are bounded operators on $L^2(A)$. Moreover

$$W_T f(x) = e^{T \cdot E_M \frac{dU_s}{ds}\big|_{s=0}} E_M f(x).$$

**Proof.**
From Lemma's 1-5 we have:

$$\lim_{N \to \infty} (E_M U_{T/N} E_M)^N f(x) = \lim_{N \to \infty} \sum_{k=0}^{\infty} \langle f | e_k \rangle \sum_{k_N=0}^{\infty} G_{k,k_N}\left(\tfrac{T}{N}\right) e_{k_N}(x) =$$

$$\lim_{N \to \infty} \sum_{k=0}^{\infty} \langle f | e_k \rangle B\left(\tfrac{T}{N}\right)^N e_k(x) = \sum_{k=0}^{\infty} \langle f | e_k \rangle e^{TB'(0)} e_k(x) =$$

$$e^{TB'(0)} E_M f(x) = e^{TE_M \frac{dU_s}{ds}\big|_{s=0}} E_M f.$$

-Let now $M$ be an arbitrary subspace of $L^2$ and $M^+$ the complement of $M$. Let $\frac{dU_s}{ds}\big|_{s=0}$ be a bounded self-adjoint operator of $L_2$ and $\{y_k(x)\}_{k \in \mathbb{N}}$ be a base of $L^2$ such that $\frac{dU_s}{ds}\big|_{s=0} y_k(x) = \lambda_k y_k(x)$, where $\lambda_k$ are eigenvalues of the above operator.

Moreover there exists bases of $M$ and $M^+$ i.e. $\{e_k(x)\}_{k \in \mathbb{N}}, \{e_k^+(x)\}_{k \in \mathbb{N}}$, respectively such that $\{y_k(x)\}_{k \in \mathbb{N}} = \{e_k\} \cup \{e_k^+\}$.

Let $y_k = \begin{cases} e_k, k \in I_1 \\ e_k^+, k \in I_2 \end{cases}$

$I_1, I_2$ are the counter sets of the elements of the bases $M$ and $M^+$, respectively.

From Main Theorem we easily get that $W_T f(x) = e^{T\left(E_M \frac{dU_s}{ds}\big|_{s=0} E_M\right)} f(x)$.

It is $f(x) = \sum_{k \in \mathbb{N}} \langle f | y_k \rangle y_k(x)$ and is easy to see someone that

$\left(E_M \frac{dU_s}{ds}\big|_{s=0} E_M\right)^n y_k(x) = \lambda_k^n y_k(x) X_{I_1}(k)$, where $X_{I_1}(k)$ is the characteristic function on $I_1$.

Thus

$$e^{TE_M \frac{dU_s}{ds}\big|_{s=0}} f(x) = \sum_{n=0}^{\infty} \left(E_M \frac{dU_s}{ds}\big|_{s=0} E_M\right)^n \frac{T^n}{n!} f(x) = \sum_{n=0}^{\infty} \frac{T^n}{n!} \sum_{k=0}^{\infty} (\lambda_k)^n X_{I_1}(k) \langle f | y_k \rangle y_k(x)$$

$$\Rightarrow W_T f(x) = W_T \Psi(x,0) = \sum_{k \in I_1} e^{\lambda_k T} \langle f | y_k \rangle y_k(x) = E_M e^{T \frac{dU_s}{ds}\big|_{s=0}} E_M \Psi(x,0)$$

So from the above method we get the next theorem:

**Theorem 1.**

The limit operator $W_T \Psi(x,0)$ can be written in the form:
$$W_T \Psi(x,0) = E_M e^{iTH} E_M \Psi(x,0),$$
Whenever $\lim_{s \to 0} \dfrac{dU_s}{ds} = iH$ and $H$ is a skew-adjoint linear operator and $U_s$ satisfies the conditions of the Main Theorem.

**Corollary.**
If $U_s = e^{sC}$, then if $C$ is self-adjoint operator and $|C| < \infty$, then
$$W_T \Psi(x,0) = E_M U_T E_M \Psi(x,0).$$

Consider now a Quantum system $Q$, whose states belong to a Hilbert space $L^2$ and whose evolution is described by the unitary operator $B$. $E_M$ is the projection that does not commute with $B$, $[E_M, B] \neq 0$, and $E_M L^2 = L_M^2$ the subspace defined by it. The initial density matrix $\rho_0$ of the system $Q$ is taken to belong to $L_M$.
$$\rho_0 = E_M \rho_0 E_M, \; Tr\rho_0 = 1.$$
The probability to find the system on $L_M$ is
$$P_M = Tr[W_T \rho_0 W_T^*].$$
Let $U_s$ is an operator with $U_0 = I$ then
$$P_M = \sum_k \langle y_k | (W_T \rho_0 W_T^*) y_k \rangle = \sum_k \langle y_k | W_T \rho_0 e^{T\lambda_k} y_k(x) X_k(I_1) \rangle =$$
$$= \sum_k \langle y_k | W_T \rho_0 e^{T\lambda_k^*} y_k(x) X_k(I_1) \rangle =$$
$$= \sum_k \langle y_k | e^{T\lambda_k^*} X_k(I_1) W_T \sum_l w_l E_{y_l(x)}(y_k(x)) \rangle =$$
$$= \sum_k \langle y_k | e^{T\lambda_k^*} X_k(I_1) W_T \delta_{k,l} w_l y_l(x) \rangle =$$
$$= \sum_k \langle y_k | e^{T\lambda_k^*} X_k(I_1) \delta_{k,l} w_l W_T(y_l(x)) \rangle =$$
$$= \sum_k \langle y_k | e^{T\lambda_k^*} X_k(I_1) \delta_{k,l} w_l e^{\lambda_l T} X_l(I_1) y_l(x) \rangle =$$
$$= \sum_{k \in I_1} \langle y_k | e^{T\lambda_k^*} w_k e^{\lambda_k T} y_k(x) \rangle =$$
$$= \sum_{k \in I_1} e^{T(\lambda_k^* + \lambda_k)} w_k$$
Thus we get the next theorem:

**Theorem 2.**
$P_M = Tr[W_T \rho_0 W_T^*] = \sum_{k \in I_1} e^{T(\lambda_k^* + \lambda_k)} w_k$, and if $\left.\dfrac{dU_s}{ds}\right|_{s=0}$ is skew adjoint then $P_M = 1$.

**Proof.**

If $\left.\dfrac{dU_s}{ds}\right|_{s=0}$ is skew adjoint then the eigenvalues $\lambda_k \in \text{Im}(\mathbb{C})$. Hence easy $P_M = 1$.

## Section 2
## Matrix Mechanics

**Definition 1**

Let $S$ be the set of all infinite-dimensional matrix, which the norm of $R = (R_{k,m})_{k,m \in \mathbb{N}} \in S$ is $\|R\| < \infty$.

**Note.** We can take as norm of $S$ the relation: $\|R\| := \sum_{k,m=0}^{\infty} |R_{k,m}| < \infty$.

The above norm must have the following properties:

$(i): \|R\| \geq 0, \ \|R\| = 0 \Leftrightarrow R = 0$

$(ii): \|R + G\| \leq \|R\| + \|G\|$

$(iii): \|aR\| = |a| \cdot \|R\|$

$(iv): \|RG\| \leq \|R\| \cdot \|G\|$

**Definition 2**

Let $R = (R_{i,j}), B = (B_{i,j})$ are two matrices in $S$ we define the mapping $J$ such that, whenever there exists the relation: $e^R = B$, $B_{i,j} = J(R_{i,j})$.

The above mapping $J$ has the following property:

Whenever: $S \ni R^{(1,2)} = (R_{i,j}^{(1,2)}), S \ni B^{(1,2)} = (B_{i,j}^{(1,2)}), e^{R^{(1,2)}} = B^{(1,2)}$ and $B^{(1)} B^{(2)} = B^{(2)} B^{(1)}$, we have:

$$\sum_k J(R_{i,k}^{(1)}) J(R_{k,j}^{(2)}) = J(R_{i,j}^{(1)} + R_{i,j}^{(2)}).$$

An explanation to this is the relation: $B^{(1)} \cdot B^{(2)} = e^{R^{(1)} + R^{(2)}}$.

We observe that the propagator $\sum_{k=0}^{\infty} R_{k,k_1}\left(\tfrac{T}{N}\right) \sum_{k_1=0}^{\infty} R_{k_1,k_2}\left(\tfrac{T}{N}\right) \cdot \ldots \cdot \sum_{k_N=0}^{\infty} R_{k_{N-1},k_N}\left(\tfrac{T}{N}\right)$ in Definition 3 of Section 1 is the $(k, k_N)$ element of some infinite dimensional matrix.

If we set $R\left(\dfrac{T}{N}\right) = \left(R_{k,m}\left(\dfrac{T}{N}\right)\right)$, then it is easy to see someone that

$$\left(R\left(\frac{T}{N}\right)\right)^N = \left(\sum_{k=0}^{\infty} R_{k,k_1}\left(\tfrac{T}{N}\right) \sum_{k_1=0}^{\infty} R_{k_1,k_2}\left(\tfrac{T}{N}\right) \cdot \ldots \cdot \sum_{k_N=0}^{\infty} R_{k_{N-1},k_N}\left(\tfrac{T}{N}\right)\right).$$

We call

$$(R'(0))_{k,m} = \lim_{s \to 0} R'_{k,m}(s) := \left\langle \left[\frac{dU_s}{ds}\right]_{s=0} e_k \,|\, e_m \right\rangle,$$

$$(R''(0))_{k,m} = \lim_{s \to 0} R''_{k,m}(s) := \left\langle \left[\frac{d^2U_s}{d^2s}\right]_{s=0} e_k \,|\, e_m \right\rangle.$$

It is easy to see someone that if $\|R'(0)\|, \|R''(0)\| < \infty$ then, for $h$, so small as we want from Taylor expansion theorem, there exists a number $\xi \in [0,s]$ such that:

$$R(h) = I + hR'(0) + \frac{h^2}{2} R''(\xi)$$

Where $I$ is the identity matrix.

**Lemma 1.**
Let $K \in S$. Then the next limit is valid with respect to norm $\|.\|$ of $S$:

$$\lim_{N \to \infty} \left\| \left(I + \tfrac{K}{N}\right)^N - e^K \right\| = 0.$$

Thus as in Section 1 $\lim_{N \to \infty} \left( R\left(\frac{T}{N}\right) \right)^N = \exp(T \cdot R'(0))$, whenever $\|R'(0)\| < \infty$.

**Theorem 1.**
Let $M$ is an arbitrary subspace of $L^2$ and let $\{e_k\}_{k=1,2,\ldots}$ is an orthonormal base of $M$. Then iff $\|R'(0)\|, \|R''(0)\| < \infty$,

$$W_T f(x) = \sum_{k=0}^{\infty} \langle f | e_k \rangle \sum_{m=0}^{\infty} J(T \cdot R'_{k,m}(0)) e_m(x)$$

When $A = (a_{k,m})$, $J(a_{k,m})$ is the mapping who gives the $(k, m)$ element of $e^A$.

**Proof.**

$$\lim_{N \to \infty} (E_M U_{T/N} E_M)^N f(x) = \lim_{N \to \infty} \sum_{k=0}^{\infty} \langle f | e_k \rangle \sum_{k_N=0}^{\infty} e_{k_N}(x) G_{k,k_N}\left(\tfrac{T}{N}\right)$$

Thus we easily get the theorem as in section 1.

## Section 3
## The semigroup and other properties

**Lemma 1.**
The operator $e^{TB'(0)}$ sends an element of $M$ into $M$.

**Proof.**
Let $f \in M$ and $e^{TB'(0)} f(x) = g(x) \in M^+ - \{0\}$, where $M^+$ is the complement of $M$.
We have that for every $\varepsilon > 0$, $\exists N_0 : \forall N > N_0$

$$\left\|e^{TB'(0)}f(x)-\sum_{k=0}^{N}\frac{T^k}{k!}(B'(0))^k f(x)\right\|_2 < \varepsilon, \text{ or } \left\|g(x)-\sum_{k=0}^{N}\frac{T^k}{k!}(B'(0))^k f(x)\right\|_2 < \varepsilon$$

Clearly $(B'(0))^k : M \to M$.

Now
$$\|g\| = \left|\left\langle g(x)-\sum_{k=0}^{N}\frac{T^k}{k!}(B'(0))^k f(x)\bigg|g\right\rangle\right| \le \left\|g(x)-\sum_{k=0}^{N}\frac{T^k}{k!}(B'(0))^k f(x)\right\| \cdot \|g\| < \varepsilon \cdot \|g\|$$

So $g = 0$, contradiction.

**Theorem 1.**

If the limit operator $W_T f(x) = \lim_{N\to\infty} (E_M U_{T/N} E_M)^N f(x)$ exists then

$$W_{T_1} W_{T_2} = W_{T_1+T_2}$$

**Proof.**

From Lemma 1

$$W_{T_1} W_{T_2} f(x) = e^{T_1 B'(0)} E_M e^{T_2 B'(0)} E_M f(x) = e^{T_1 B'(0)} e^{T_2 B'(0)} E_M f(x) =$$
$$e^{(T_1+T_2)B'(0)} E_M f(x) = W_{T_1+T_2} f(x).$$

**Lemma 2.**

Whenever $M$ is dense in $L_2(\mathbb{R})$, $W_T f(x) = E_M e^{T\frac{dU_s}{ds}\big|_{s=0}} E_M f(x)$.

**Proof.**

Let $M$ has base $\{e_k\}_{k\in\mathbb{N}}$.

We write $W_T f(x) = E_M e^{T\cdot E_M \frac{dU_s}{ds}\big|_{s=0} E_M} E_M f(x)$. From **[M,S]**, there exists a self-adjoint operator $C$ such that $W_T f(x) = E_M e^{iTC} E_M f(x)$

It is clear from the main theorem that for every $f \in L^2(\mathbb{R})$ we have the relation (see **[M,S]**): $e^{TB'(0)} E_M f(x) = E_M e^{iTC} E_M f(x) \Leftrightarrow$

$$e^{TB'(0)} \sum_{k=0}^{\infty} \langle f|e_k\rangle e_k(x) = E_M e^{iTC} \sum_{k=0}^{\infty} \langle f|e_k\rangle e_k(x) \Leftrightarrow$$

$$\sum_{k=0}^{\infty} \langle f|e_k\rangle e^{TB'(0)} e_k(x) = \sum_{k=0}^{\infty} \langle f|e_k\rangle E_M e^{iTC} e_k(x) \Leftrightarrow$$

(a): $\sum_{k=0}^{\infty} \langle f|e_k\rangle e^{TB'(0)} e_k(x) = \sum_{k=0}^{\infty} \langle f|e_k\rangle \sum_{m=0}^{\infty} \langle e^{iTC} e_k(x)|e_m(x)\rangle e_m(x)$

For $f(x) = e_j(x)$ in (a) we get

$$e^{TB'(0)} e_j(x) = \sum_{m=0}^{\infty} \langle e^{iTC} e_j(x)|e_m(x)\rangle e_m(x)$$

Now we expand the exponentials in Taylor series. (The operators $B'(0)$ and $C$ are bounded.)

$$\sum_{r=0}^{\infty}\frac{T^r}{r!}(B'(0))^r(e_j(x)) = \sum_{m=0}^{\infty}\left\langle\sum_{r=0}^{\infty}\frac{T^r}{r!}(iC)^r(e_j(x))\bigg|e_m(x)\right\rangle e_m(x) \Leftrightarrow$$

$$\sum_{r=0}^{\infty}\frac{T^r}{r!}(B'(0))^r(e_j(x)) = \sum_{r=0}^{\infty}\frac{T^r}{r!}\sum_{m=0}^{\infty}\left\langle(iC)^r(e_j(x))\big|e_m(x)\right\rangle e_m(x)$$

Since the relation above exists for every $T > 0$ we get:

$$(B'(0))^r(e_j(x)) = \sum_{m=-\infty}^{\infty}\left\langle(iC)^r(e_j(x))\big|e_m(x)\right\rangle e_m(x), \text{ for every } r = 0,1,2,\ldots$$

Thus for $r = 1$ we get

$$B'(0)(e_j(x)) = \sum_{m=0}^{\infty}\left\langle iC(e_j(x))\big|e_m(x)\right\rangle e_m(x) \text{ or equivalently}$$

$$\sum_{m=0}^{\infty}R'_{k,m}(0)e_m(x) = \sum_{m=-\infty}^{\infty}\left\langle iC(e_j(x))\big|e_m(x)\right\rangle e_m(x) \text{ or from the definition of the } R'$$

$$\sum_{m=-\infty}^{\infty}\left\langle\left(\frac{dU_s}{ds}\bigg|_{s=0} - iC\right)e_j(x)\bigg|e_m(x)\right\rangle e_m(x) = 0$$

Let now $G = \dfrac{dU_s}{ds}\bigg|_{s=0} - iC$. We will show that $G = 0$ in $L^2(A)$.

We have $\sum_{m=0}^{\infty}\left\langle Ge_j(x)\big|e_m(x)\right\rangle e_m(x) = 0$ for every $j$ in $\mathbb{Z}$ or

$\left\langle Ge_j(x)\big|e_m(x)\right\rangle = 0$ for every $j$ in $\mathbb{Z}$.

From the last relation it follows that $\left\langle Gf\big|g\right\rangle = 0$ whenever $f, g \in M$.

From the density of $M$ in $L^2(\mathbb{R})$ we get that for every $h \in L^2(A)$ there exists a sequence $g_k(x) \in M$, $k=1,2,3,\ldots$ such that $\forall \varepsilon > 0$, $\exists N > N_0 : \left\|h(x) - \sum_{k=1}^{N}g_k(x)\right\|_2 < \varepsilon$.

Set $O_N := \left\langle G\left(h(x) - \sum_{k=1}^{N}c_k g_k(x)\right)\bigg|g\right\rangle$, $g$ in $M$

$$O_N = \left\langle Gh(x)\big|g\right\rangle - \sum_{k=1}^{N}c_k\left\langle Gg_k(x)\big|g\right\rangle = \left\langle Gh(x)\big|g\right\rangle$$

But $\left|\left\langle G\left(h(x) - \sum_{k=1}^{N}c_k g_k(x)\right)\bigg|g\right\rangle\right| < |G|\cdot\|g\|_2\,\varepsilon$, where $|G|$ is the norm of $G$.

So $\left\langle Gh(x)\big|g\right\rangle = 0$ for every $h \in L^2(A)$, $g$ in $M$.

Let now $h^{(2)} \in L^2(A)$, because $M$ is dense in $L^2(A)$, $\forall \varepsilon^{(2)} > 0$, $\exists N > N_0$:

$$\left\|h^{(2)}(x) - \sum_{k=1}^{N}g_k^{(2)}(x)\right\|_2 < \varepsilon^{(2)}.$$

$$\left|\left\langle Gh(x)\middle|h^{(2)}(x)-\sum_{k=1}^{N}c_k g_k(x)\right\rangle\right|=\left|\left\langle Gh(x)\middle|h^{(2)}(x)\right\rangle\right|$$

Also

$$\left|\left\langle Gh(x)\middle|h^{(2)}(x)-\sum_{k=1}^{N}c_k g_k(x)\right\rangle\right|\leq\|Gh(x)\|_2\left\|h^{(2)}(x)-\sum_{k=1}^{N}c_k g_k(x)\right\|_2<\varepsilon^{(2)}|G|\cdot\|h^{(2)}\|_2$$

Thus

$$\left|\left\langle Gh(x)\middle|h^{(2)}(x)\right\rangle\right|<\varepsilon^{(2)}|G|\cdot\|h^{(2)}\|_2,\text{ and }G=0\text{ in }L^2(A)\text{ so }\left.\frac{dU_s}{ds}\right|_{s=0}=iC.$$

Next we examine the case of the operator

$$W_T:=\lim_{N\to\infty}\prod_{k=1}^{N}\left(E_M U_{t_k} E_M\right)$$

Where $E_M$ is projection in the space $M\subseteq L^2(A)$ and $U_s=e^{isH}$ is a unitary operator (H is skew adjoint). The $t_k$ form a sequence such that $\sum_{m=1}^{\infty}t_m=T<\infty$.

We will show that

**Theorem 2.**
$$W_T=E_M e^{T\cdot H}E_M$$

**Proof.**
Let $\{y_k\}$, $\{\lambda_k\}$, $k\in\mathbb{N}$ are the eigenvectors and eigenvalues of $H$ respectively. It is clear that if $\{y_k\}=\{e_k\}\cup\{e_k^+\}$ with $e_k$ is a base of $M$ $e_k^+$ is a base of $M^+$ and $I_1,I_2$ are the counter of the elements of the two subspaces $M$, $M^+$, We can compute the operator:

$$W_T^N:=\prod_{k=1}^{N}\left(E_M U_{t_k} E_M\right).$$

It is: $W_T^1 f(x)=\prod_{k=1}^{1}\left(E_M U_{t_k} E_M f(x)\right)=E_M\sum_{k=-\infty}^{\infty}\langle E_M f|e_k\rangle e^{t_1 H}(y_k(x))=$

$$\sum_{k=-\infty}^{\infty}\langle E_M f|e_k\rangle\sum_{m=0}^{\infty}\frac{t_1^m}{m!}\lambda_k^m X_{I_1}(k)y_k(x)=\sum_{k=-\infty}^{\infty}\langle E_M f|y_k\rangle e^{t_1\lambda_k}X_{I_1}(k)y_k(x)$$

In the same way we get:

$$W_T^2 f(x)=\prod_{k=1}^{2}\left(E_M U_{t_k} E_M f(x)\right)=\sum_{k=-\infty}^{\infty}\langle E_M f|y_k\rangle e^{(t_1+t_2)\lambda_k}\left(X_{I_1}(k)\right)^2 y_k(x).\text{ And}$$

easily we deduce

$$W_T^N := \prod_{k=1}^{N}(E_M U_{t_k} E_M) = \sum_{k=-\infty}^{\infty} \langle E_M f | y_k \rangle e^{\left(\sum_{m=1}^{N} t_m\right)\lambda_k} (X_{I_1}(k))^N y_k(x).$$ Taking the limit $N \to \infty$ we get $$W_T := \prod_{k=1}^{\infty}(E_M U_{t_k} E_M) = \sum_{k=-\infty}^{\infty} \langle E_M f | y_k \rangle e^{T\lambda_k} X_{I_1}(k) y_k(x).$$

From another point of view we have

$$E_M e^{TH} E_M f(x) = \ldots = \sum_{k=-\infty}^{\infty} \langle E_M f | y_k \rangle e^{T\lambda_k} X_{I_1}(k) y_k(x)$$ and we complete the proof.

An extension of the semigroup property is the following theorem.

**Theorem 3.**

If we observe the particle in random times $t_{1,k}, t_{2,k}$ with $T_1 = \sum_{k=1}^{\infty} t_{1,k}$, $T_2 = \sum_{k=1}^{\infty} t_{2,k}$ then

$$W_{T_1} W_{T_2} = W_{T_1+T_2}$$

**Proof**
Easy

**Theorem 4.**

It is easy to recover $W_T$ when we now the values $W_T(e^{ixa})$, $a \in \mathbb{R}$. In special the next formulation is valid:

$$W_T f(x) = \frac{1}{2\pi} \langle E_M f^\wedge(.) | W_T(e^{ix(.)}) \rangle$$

**Proof.**
Using Parseval`s formula we get

$$W_T \Psi_0 = \frac{1}{2\pi} \langle E_M \Psi_0^\wedge(a) | W_T(e^{ita}) \rangle =$$

$$\langle E_M \Psi_0(t) | W_T(\delta(t-a))(x) \rangle =$$

$$\left\langle E_M \Psi_0(t) \middle| \sum_{k=-\infty}^{\infty} \langle \delta(t-a) | y_k(t) \rangle e^{T\lambda_k} X_{I_1}(k) y_k(x) \right\rangle =$$

$$\left\langle E_M \Psi_0(t) \middle| \sum_{k=-\infty}^{\infty} \overline{y_k(t)} e^{T\lambda_k} X_{I_1}(k) y_k(x) \right\rangle =$$

$$\sum_{k=-\infty}^{\infty} \left( \int_R E_M \Psi_0(a) \overline{y_k(a)} da \right) e^{T\lambda_k} X_{I_1}(k) y_k(x) =$$

$$\sum_{k=-\infty}^{\infty} \langle E_M \Psi_0 | y_k \rangle e^{T\lambda_k} X_{I_1}(k) y_k(x).$$

**Theorem 5.**
If $f$ is some entire function and $F$ skew adjoint operator then $[f(F), W_T] = 0$ in $M$.

**Proof.**
Easy

## Section 4
## Applications and some theorems on Zenon effect

**Definition 1.**
We define $A$ to be the set of all differentiable functions $\Theta(x,y): \mathbb{R} \times \mathbb{R} \to \mathbb{R}$ and $\Theta(x,0) = 1$.

**Definition 2.**
We define the operator $F_s : L^2(\mathbb{R}) \to L^2(\mathbb{R})$, $F_s f(x) = \int_{-\infty}^{\infty} f(t) g_s(x-t) dt$, with

$g_s{}^\wedge(\gamma) = e^{i\Theta(\gamma,s)}$, and $\Theta(x,y) \in A$.

**Lemma 1.**
The operator $F_S$ defined above is unitary.

**Proof.**
First we observe that $F_S$ is an isometry, in fact we apply Parseval`s identity to get:

$$\|F_s f(x)\|_2 = \|(F_s f)^\wedge(\gamma)\|_2 = \|f^\wedge(\gamma) e^{i\Theta(\gamma,s)}\|_2 = \|f^\wedge(\gamma)\|_2 = \|f\|_2$$

Next we have to show that there exist $F_s^{-1}$ such that $F_s^{-1} F_s = I$, $I$ is the identity operator.

We have easily:
$(F_s f)^\wedge(\gamma) = f(t)^\wedge(\gamma) g_s{}^\wedge(\gamma) = f^\wedge(\gamma) e^{i\Theta(\gamma,s)}$
thus
$f^\wedge(\gamma) = (F_s f)^\wedge(\gamma) e^{-i\Theta(\gamma,s)}$
and so $f^\wedge(\gamma) = (F_s f)^\wedge(\gamma) e^{-i\Theta(\gamma,s)}$.

We apply in the last expression the Fourier inversion formula:
$$f(x) = \frac{1}{2\pi} \int_{-\infty}^{\infty} f^\wedge(\gamma) e^{i\gamma x} d\gamma = \frac{1}{2\pi} \int_{-\infty}^{\infty} (F_s f)^\wedge(\gamma) e^{-i\Theta(\gamma,s)} e^{i\gamma x} d\gamma$$

$$F_s^{-1} h(x) = \frac{1}{2\pi} \int_{-\infty}^{\infty} \int_{-\infty}^{\infty} h(t) e^{-it\gamma} dt \cdot e^{-i\Theta(\gamma,s)} e^{i\gamma x} d\gamma$$

Therefore there exist $F_s^{-1}: L^2(\mathbb{R}) \to L^2(\mathbb{R})$, and $h \to F_s^{-1} h$ is defined as follows:
And this completes the proof.

**Theorem 1.**

If $U_s f(x) = F_s f(x) = \int_{\mathbb{R}} f(t) g_s(x-t) dt$ as above, then:

$$W_T \Psi(x,0) = \frac{1}{2\pi} \int_{\mathbb{R}} (E_M \Psi)^\wedge (\gamma, 0) e^{Ti \left[\frac{\partial \Theta(\gamma,s)}{\partial s}\right]_{s=0}} e^{i\gamma x} d\gamma$$

**Proof.**

$F_s f^\wedge(\gamma) = f^\wedge(\gamma) g_s^\wedge(\gamma) \Rightarrow F_s^n f^\wedge(\gamma) = f^\wedge(\gamma)(g_s^\wedge(\gamma))^n \Rightarrow$

$$e^{T \cdot \frac{d}{ds} F}\bigg|_{s=0} = \sum_{n=0}^{\infty} \frac{T^n}{n!} f^\wedge(\gamma)(g_s(\gamma))^n = f^\wedge(\gamma) e^{T g_s^\wedge(\gamma)}.$$

Thus using the inverse Fourier Transform we get

$$e^{T \cdot \frac{d}{ds} F_s}\bigg|_{s=0} f(x) = \frac{1}{2\pi} \int_{\mathbb{R}} f^\wedge(\gamma) e^{T \frac{d}{ds} g_s^\wedge(\gamma)\big|_{s=0}} e^{ix\gamma} d\gamma$$

**Theorem 2.**

If $U_s f(x) = e^{isx} f(x)$ then $W_T \Psi(x,0) = E_M \Psi(x,0) e^{iTx}$

**Application 1.**

Let $F_s$ an operator and $\Theta$ in A such that: $\Theta(\gamma, s) = s \cdot c + s^2 \phi(\gamma, s)$, with

$:\exists \left[\frac{\partial}{\partial s} \phi(\gamma, s)\right]_{s=0}$ and $\left|\left[\frac{\partial}{\partial s} \phi(\gamma, s)\right]_{s=0}\right| < \infty$, for every $\gamma$ in $\mathbb{R}$, c-constant and M

dense in $L_2(\mathbb{R})$. Then the limit operator has the form.

$$W_T^\infty f(x) = e^{ciT} E_M f(x)$$

**Proof.**

It is $\left.\frac{\partial \Theta(\gamma, s)}{\partial s}\right|_{s=0} = c$, for every $\gamma \in \mathbb{R}$. Thus

$$W_T f(x) = \frac{1}{2\pi} \int_{-\infty}^{\infty} E_M f^\wedge(\gamma) \exp\left(iT \left.\frac{\partial \Theta(\gamma, s)}{\partial s}\right|_{s=0}\right) e^{i\gamma x} d\gamma =$$

$$\frac{1}{2\pi} \int_{-\infty}^{\infty} E_M f^\wedge(\gamma) \exp(iTc) e^{i\gamma x} d\gamma = e^{cTi} E_M f(x)$$

**Note.**

(Let $F_S$ be a unitary operator such that: $F_{T/N} f(x) = \frac{N \cdot i}{2\sqrt{\pi T}} e^{\frac{iT}{N}} \int_{-\infty}^{\infty} f(t) e^{\frac{-i(x-t)^2}{4T^2} N^2} dt$.

Then the limit operator of Zenon effect exists and is well defined i.e.
$W_T^\infty f(x) = e^{iT} E_M f(x)$, This is the case of the above application with $c = 1$ and $\phi(\gamma, s) = \gamma^2$).

**Application 2.**
Let $\Delta = [0, L]$ and the projection in $L_2(\Delta)$ is $E_\Delta f(x) = X_\Delta(x) f(x)$, $X_\Delta$ is the characteristic function on $\Delta$. Let $F_s$ be a unitary operator such that:
$F_s f(x) = f(x+s)$ so $g_s{}^\wedge(\gamma) = e^{i\gamma \cdot s}$ and

$$W_T f(x) = \frac{1}{2\pi} \int_{-\infty}^{\infty} E_M f^\wedge(\gamma) e^{i\gamma x + iT \frac{\partial \Theta(\gamma, s)}{\partial s}\big|_{s=0}} d\gamma = \frac{1}{2\pi} \int_{-\infty}^{\infty} E_M f^\wedge(\gamma) e^{i\gamma x + iT\gamma} d\gamma = E_M f(x+T).$$

**Application 3.**

If we take for $U_s = \sum_{k=0}^{\infty} a_k(s) \frac{d^k}{dx^k}$, with $0 = a_1(0) = a_2(0) = \ldots$, $a_0(0) = 1$ and then set $a_k'(0) = b_k$, $\frac{d}{ds} U_s \big|_{s=0} = \sum_k b_k \frac{d^k}{ds^k}$. One can easily see that

$$W_T \Psi(x, 0) = \frac{1}{2\pi} \int_{-\infty}^{\infty} (E_M \Psi)^\wedge(\gamma, 0) e^{T \sum_{k=0}^{\infty} b_k \gamma^k} e^{i\gamma x} d\gamma$$

**Section 5**
**Zenon effect in linear non unitary time evolution**
**(another approx)**

From the above calculations it follows that if we have a Quantum system whose evolution is governed by a semigroup low i.e. $\Psi(x, t) = e^{tB} \Psi(x, 0)$, then the survival probability after continuous observation will be
$$P(T) = e^{T \operatorname{Re}(\langle B_{\Psi_0} \rangle)}$$
It is clear that if we have to deal with a system with $B = \frac{-i}{\hbar} H$ then due to self adjoint of $H$ the probability becomes 1.

Let us now consider a particle moving in the real line with mass $m$ and charge $q$ in an electromagnetic field described by an electrical potential $V(x)$ and another potential $A(x)$.

The equation who describes the above system is given by
$$i\hbar \frac{\partial \Psi(x, t)}{\partial t} = \left[ \frac{1}{2m} \left( \frac{\hbar}{i} \frac{\partial}{\partial x} - \frac{q}{c} A(x) \right)^2 + qV(x) \right] \Psi(x, t)$$

We assume that the motion is governed by a semigroup low whose generator is $B$ and the probability to find the particle not decaying is as above $P(T) = e^{T \text{Re}(<B_{\Psi_0}>)}$
For these assumptions after some calculations we find that the survival probability is

$$P(T) = e^{\frac{Tq}{mc} \text{Re}\left(\int_R \Psi_0^* \left(A(x)\frac{\partial \Psi_0}{\partial x} + \frac{1}{2}\Psi_0 \frac{\partial A(x)}{\partial x}\right)dx\right)}$$

As someone can see the above probability is independed of the existence of the potential $V(x)$.

Let us proceed with an example

Let $\Psi(x,0) = \sqrt[4]{\frac{2}{\pi}} e^{-x^2}$ is the initial state of the particle. The potential $A$ is given by $A(x) = ax^s, s \in \mathbb{R}, a > 0$. Then for these assumptions we get that the survival probability is

$$P(T) = e^{\frac{Tqa}{mc\pi} 2^{-s/2}(\cos(\pi s)-1)\Gamma\left(1+\frac{s}{2}\right)}$$

The **exact** equation Re(<B>) = 0, for this case have solutions $s=2k$, $k$-integer.
For $s \in \{2k : k \in \mathbb{Z}\}$ we get the valid values of $s$.

So we make the conclusion that if (1): $A(x) = \sum_{k=0}^{\infty} a_{2k} x^{2k}, a_{2k} \in \mathbb{R}$, and if $\Psi(x,0) = f(x)$, with even, or odd $f \in L^2(\mathbb{R}) \cap \mathbb{R}$, differentiable in $\mathbb{R}$, then: $P = 1$ **always**. (If exist such potentials)**.**

## Section 6
## Zenon effect in Relativistic Quantum Mechanics

The Klein-Gordon equation for a particle moving without restrictions in a real line read as

$$\frac{\partial^2 \Psi(x,t)}{\partial x^2} = \frac{1}{c^2}\frac{\partial^2 \Psi(x,t)}{\partial t^2} + \gamma_0^2 \Psi(x,t): \quad \text{(K-G)}$$

where $\gamma_0 = \frac{mc}{\hbar}$, $m$ is the mass of the particle $c$ is the velocity of light and $\hbar$ is the Planck constant see **[G]**.
We first solve (K-G), then we find an operator $A$ such that $e^{tA}\Psi(x,0) = \Psi(x,T)$, where $\Psi(x,0) = f(x)$ is the initial condition of the system. Then we find the conditions for the occurrence of the Zenon effect.

We take the Fourier transform with respect to $x$ in (K-G) to obtain

$$\gamma^2 F(\Psi)(\gamma,t) = \frac{1}{c^2}\frac{\partial^2 F(\Psi)(\gamma,t)}{\partial t^2} + \gamma_0^2 F(\Psi)(\gamma,t).$$

We solve the above equation with respect to $t$ $(F(\Psi)(\gamma,t) = \int_R \Psi(x,t)e^{-ix\gamma}dx)$

There exists functions $c_1{}^\wedge(\gamma)$, $c_2{}^\wedge(\gamma)$, such that:

$F(\Psi)(\gamma,t) = c_1{}^\wedge(\gamma)e^{tc\sqrt{\gamma^2-\gamma_0^2}} + c_2{}^\wedge(\gamma)e^{-tc\sqrt{\gamma^2-\gamma_0^2}}$. Thus if we inverse the Fourier transform we get the solution of (K-G).

$\Psi(x,t) = \dfrac{1}{2\pi}\int_R F(\Psi(\gamma,t))e^{ix\gamma}d\gamma \Leftrightarrow$

$(S0): \Psi(x,t) = \dfrac{1}{2\pi}\int_R \left(c_1{}^\wedge(\gamma)e^{tc\sqrt{\gamma^2-\gamma_0^2}} + c_2{}^\wedge(\gamma)e^{-tc\sqrt{\gamma^2-\gamma_0^2}}\right)e^{ix\gamma}d\gamma$

In order to have convergence we take $c_1{}^\wedge(\gamma) = 0$. So for $t = 0$ we get

$\Psi(x,0) = \dfrac{1}{2\pi}\int_R c_2{}^\wedge(\gamma)e^{ix\gamma}d\gamma = c_2(x) = f(x)$. Thus the solution of (K-G) equation with conditions $\Psi(x,0) = f(x)$ and $\lim_{t\to\infty}\Psi(x,t) = 0$, is

**(S1):** $\Psi(x,t) = \Lambda \cdot \int_R f^\wedge(\gamma)e^{-tc\sqrt{\gamma^2-\gamma_0^2}+ix\gamma}d\gamma$, for some constant of normalization $\Lambda$.

Now we write equation **(S1)** in the form

$\Psi(x,t) = \Lambda\int_R f^\wedge(\gamma)\sum_{k=0}^{\infty}\dfrac{1}{k!}\left(-tc\sqrt{\gamma^2-\gamma_0^2}\right)^k e^{ix\gamma}d\gamma =$

$\sum_{k=0}^{\infty}\dfrac{1}{k!}\Lambda\int_R f^\wedge(\gamma)\left(-tc\sqrt{\gamma^2-\gamma_0^2}\right)^k e^{ix\gamma}d\gamma$. From this equation we get that if $B$ is an operator such that $Bf(x) = -c\Lambda \cdot \int_R f^\wedge(\gamma)\sqrt{\gamma^2-\gamma_0^2}\,e^{ix\gamma}d\gamma$ then, $e^{tB}\Psi(x,0) = \Psi(x,t)$.

The survival probability is $P_\tau = |\langle\Psi(x,0)|\Psi(x,\tau)\rangle|^2$.

For small $\tau$: $e^{\tau B} = 1 + \tau B + \tfrac{1}{2}\tau B^2$.

Thus for small $\tau$: $P_\tau = \left|\langle\Psi(x,0)|(1+\tau B+\tfrac{1}{2}\tau B^2)\Psi(x,0)\rangle\right|^2 =$

$1 + \tau(\langle\Psi(x,0)|B\Psi(x,0)\rangle + \langle B\Psi(x,0)|\Psi(x,0)\rangle) + O(\tau^2)$
for $\tau = T/N$

$P_{T/N} = 1 + \dfrac{T}{N}(\langle\Psi(x,0)|B\Psi(x,0)\rangle + \langle B\Psi(x,0)|\Psi(x,0)\rangle) + O\left(\left(\dfrac{T}{N}\right)^2\right)$.

$T$ is any finite time interval and $N$ the number of observations during $T$. If the system is $n$-times observed during $T$ the survival probability is

$P_N = \left(1 + \dfrac{T}{N}(\langle\Psi(x,0)|A\Psi(x,0)\rangle + \langle A\Psi(x,0)|\Psi(x,0)\rangle) + O\left(\left(\dfrac{T}{N}\right)^2\right)\right)^N$, taking now the

limit with respect to $N$,

$$P = \lim_{N \to \infty} P_N = e^{T(\langle \Psi(x,0)|B\Psi(x,0)\rangle + \langle B\Psi(x,0)|\Psi(x,0)\rangle)}$$

but

$$\langle \Psi(x,0)|A\Psi(x,0)\rangle + \langle A\Psi(x,0)|\Psi(x,0)\rangle = \int_R \text{Re}\left(\overline{\Psi(x,0)}A\Psi(x,0)\right)dx =$$

$$\int_{\mathbb{R}} \text{Re}\left(\overline{\Psi(x,0)} \cdot (-c\Lambda)\int_{\mathbb{R}} \Psi^\wedge(\gamma,0)\sqrt{\gamma^2 - \gamma_0^2}\, e^{ix\gamma}d\gamma\right)dx =$$

$$-c\Lambda \, \text{Re}\left(\int_{\mathbb{R}} \Psi^\wedge(\gamma,0)\sqrt{\gamma^2 - \gamma_0^2}\left(\int_R \overline{\Psi(x,0)}e^{ix\gamma}dx\right)d\gamma\right) =$$

$$-c\Lambda \, \text{Re}\left(\int_{\mathbb{R}} |\Psi^\wedge(\gamma,0)|^2 \cdot \sqrt{\gamma^2 - \gamma_0^2}\, d\gamma\right).$$

If $X_\Omega(x)$ is the characteristic function in $\Omega = (-\infty, -\gamma_0] \cup [\gamma_0, +\infty)$
we get that the Zenon effect is occur if and only if

$\int_{\mathbb{R}} |\Psi^\wedge(\gamma,0)|^2 \cdot \sqrt{\gamma^2 - \gamma_0^2}\, X_\Omega(\gamma)d\gamma = 0$. But as someone can see this is not always true.

The quantity $Q(\gamma_0) := -c\Lambda \int_R |\Psi^\wedge(\gamma,0)|^2 \cdot \sqrt{\gamma^2 - \gamma_0^2}\, X_\Omega(\gamma)d\gamma$ may differ from 0.

Let us see what happens.

Let $\Lambda = 1$, $\Psi(x,0) = \frac{1}{2}e^{-|x|}$, be the initial state of the system. The constant $\gamma_0$ is $\frac{mc}{\hbar}$. If the particle is an electron then $m = 9.108 \times 10^{-28}$ and $\gamma_0 = 2.6 \times 10^{10}$. With these conditions we find $Q(\gamma_0) \cong -3 \times 10^{-11}$. This is very small, thus the probability $P$ is very near to 1, but not 1. There is a very small probability that the particle may move. The non 1 probability can be very near to 1.

If take as initial condition in (K-G) equation the Gaussian i.e. $\Psi(x,0) = \frac{1}{\sqrt{\pi}}e^{-x^2}$, then

$\Psi^\wedge(\gamma,0) = e^{-\gamma^2/4}$, $Q(100) \cong -7.5 \times 10^{-2153}$, ($m = 3.5 \times 10^{-36}$).

We introduce next a series of initial conditions such that the above quantity $Q$ is always 0.

It is obvious that $Q = 0$ if and only if $\Psi^\wedge(\gamma,0) = f^\wedge(\gamma) = 0$, for $w = |\gamma_0| \leq |\gamma|$. These functions *f* are called Band Limited. There is a theorem of Shannon **[Pa]** who says that every Band Limited function $f(x)$ can be written in the form

$$f(x) = \sum_{k \in Z} f\left(\frac{k\pi}{w}\right)\frac{\sin(wx - k\pi)}{wx - k\pi}.$$

Thus speaking mathematically the **Q.Z.E**. occur for all times *T*, if and only if the function is Band limited in [-*w*, *w*].

The time of observing the particle also plays essential role in the occurrence of Zenon effect.
If for example we observe an electron for times up to $10^{10}$ sec and the initial state of the electron is not Band Limited, then the probability not to change state yould be $P = e^{TQ(\gamma_0)} \cong 0.74$.

Next we examine whether or not the state function $\Psi(x,t)$ is band limited.

Let $H$ is the Hamiltonian of a Quantum system, $H$ obviously is self adjoint. Let $\{y_s(x)\}_{s\in\mathbb{Z}}$, are the eigenvectors of $H$ and $\{\lambda_s\}_{s\in\mathbb{Z}}$ its eigenvalues.

Then $\Psi(x,0) = \sum_{k=-\infty}^{\infty} \langle \Psi_0 | y_k \rangle y_k(x) \Rightarrow \Psi(x,t) = \sum_{k=-\infty}^{\infty} e^{\lambda_k t} \langle \Psi_0 | y_k \rangle y_k(x)$.

Taking now the Fourier Transform we get $\Psi\wedge(\gamma,t) = \sum_{k=-\infty}^{\infty} e^{\lambda_k t} \langle \Psi_0 | y_k \rangle y_k\wedge(\gamma)$.

$\Psi\wedge(\gamma,0) = 0, \Psi\wedge(\gamma,t) = 0, |\gamma| > |\gamma_0| \Leftrightarrow y_k\wedge(\gamma) = 0, |\gamma| > |\gamma_0|$, for some constant $\gamma_0$ (if $\{y_s(x)\}_{\in\mathbb{Z}}$, is a base of $L^2(\mathbb{R})$ then $\{y_s\wedge(\gamma)\}_{s\in\mathbb{Z}}$ is also a base).

Now according to Shannon theorem (without the lost of generality), we may assume that $S_k(x) = \dfrac{\sin(x-k)}{x-k}$: $y_s(x) = \sum_{k\in\mathbb{Z}} y_s(k) \cdot S_k(x)$ and $\gamma_0 = 1$.

But $\delta_{s,j} = \langle y_s(x) | y_j(x) \rangle = \sum_{k\in\mathbb{Z}} y_s(k) \langle S_k(x) | y_j(x) \rangle = \sum_{k\in\mathbb{Z}} y_s(k) \dfrac{1}{2\pi} \langle S_k\wedge(\gamma) | y_j\wedge(\gamma) \rangle$.

Hence we have: $\dfrac{1}{2\pi} \sum_{k\in\mathbb{Z}} y_s(k) \int_{-1}^{1} y_j\wedge(\gamma) d\gamma = 0$, for $s \neq j$ or $\int_{\mathbb{R}} y_j\wedge(\gamma) d\gamma = 0$, or

$\int_{-\infty}^{\infty} f\wedge(\gamma) d\gamma = 0$ for every $f$, which is a contradiction.

Summarizing the above we get the next Theorem

**Theorem.**
The functions that describe the state of a particle in Relativistic Quantum Mechanics are not Band Limited.

Thus we can say that if we observe a particle, an electron say then after a very large period of time according Klein's-Gordon equation the particle will change state.

# References.